\documentclass[a4paper,usenatbib]{mnras}

\usepackage{newtxtext,newtxmath}
\usepackage[T1]{fontenc}
\usepackage{ae,aecompl}

\usepackage{graphicx}	% Including figure files
\usepackage{amsmath}	% Advanced maths commands
\usepackage{amssymb}	% Extra maths symbols

\newcommand{\f}{{\cal F}}
\newcommand{\g}{{\cal G}}

\title[Shape, Color, and Distance in Weak Flexion]{Shape, Color, and Distance in Weak Gravitational Flexion}

\author[Fabritius, Arena, Goldberg]{
Joseph M. Fabritius II,$^{1}$\thanks{E-mail: joseph.m.fabritius@drexel.edu}
Evan J. Arena,$^{1}$
David M. Goldberg$^{1}$
\\
$^{1}$Department of Physics, Drexel University, Philadelphia, PA 19104\\
}

\date{Accepted XXX. Received YYY; in original form ZZZ}

\pubyear{2020}

\begin{document}
\label{firstpage}
\pagerange{\pageref{firstpage}--\pageref{lastpage}}
\maketitle

\begin{abstract}
Canonically, elliptical galaxies might be expected to have a perfect rotational symmetry, making them ideal targets for flexion studies — however, this assumption hasn’t been tested. We have undertaken an analysis of low and high redshift galaxy catalogs of known morphological type with a new gravitational lensing code, \texttt{Lenser}. Using color measurements in the $u-r$ bands and fit S{\'e}rsic index values, objects with characteristics consistent with \textit{early}-type galaxies are found to have a lower intrinsic scatter in flexion signal than \textit{late}-type galaxies. We find this measured flexion noise can be reduced by more than a factor of two at both low and high redshift.
\end{abstract}


%%%%%%%%%%%%%%%%%%%%%%%%%%%%%%%%%%%%%%%%%%%%%%%%%%

%%%%%%%%%%%%%%%%% BODY OF PAPER %%%%%%%%%%%%%%%%%%

\section{Introduction}\label{sec:Introduction}

Over the past several decades, gravitational lensing has proven an invaluable tool in direct measurement of Dark Matter structure in clusters. For a review see \citet{schneider20165} and references therein. The observation of multiple strongly lensed images can be used to determine the radial profile of galaxy clusters yet proves insufficient in probing areas of less-prominent overdensity in the underlying cluster field. Turning to higher-order distortions, such as \textit{flexion}, provides a method to help identify these low-mass signals \citep{Bacon, Er/Schneider,Goldberg/Bacon,Lasky/Fluke}. Typical lensing analyses use shear by itself, but given that shear and flexion operate on different scales flexion provides an additional avenue for constraining lensing-based measurements in large galaxy clusters. 

Understanding the distribution of intrinsic flexion and its behavior with respect to a source galaxy's shape and color should greatly improve flexion-based techniques. The intrinsic distribution in the flexion signal of a set of source objects will factor into the noise threshold in determining a likely induced flexion signal from lensing fields.

\section{Background}\label{sec:Background}

\subsection{Flexion Formalism}\label{sec:Flexion_Formalism}

Using the thin lens model, we relate the convergence to a dimensionless potential with $\nabla^{2}\psi$ = $2 \kappa$. The coordinate backwards mapping problem from foreground positions ($\vec{\theta}$) to background ($\vec{\beta}$) is then related via the linear relation of this potential:

\begin{equation}\label{eq:lensTaylor}
\beta_i = \delta_{ij}\theta^j - \psi_{,ij}\theta^{j} - \frac{1}{2}\psi_{,ijk}\theta^{j}\theta^{k}
\end{equation}

\noindent and the indices vary over the \textit{x} and \textit{y} cardinal measurements. We define the complex derivative operator $\partial$ = $\partial_{1}$+ $i\partial_{2}$. The lensing tensors in the expansion may then be related to the observed lensing effects via

\begin{align}
\f =& |\f|e^{i\phi} = \frac{1}{2}\partial\partial^{*}\partial\psi = \partial\kappa\\
\g =& |\g|e^{3i\phi} = \frac{1}{2}\partial\partial\partial\psi = \partial\gamma
\end{align}

\noindent where we note the use of complex notation shows $\f$ as gradient-like.

The first flexion, ${\cal F}$ presents itself through a centroid shift in the lensed object.  The
second flexion, ${\cal G}$, is a bit tougher to visualize, as it is a
triangular field distortion with a three-fold rotational symmetry. The present study focuses on the measurement of ${\cal F}$ exclusively, as first flexion provides a more robust measure for readily identifying observed effects in lensing systems. 

A major hurdle in flexion studies is understanding the noise associated with a measured signal. While canonically, elliptical galaxies have concentric elliptical isophotes, deviations from this will contribute to a measurement of the flexion, even when there are no lensing fields present. As galaxies are primarily classifiable as elliptical (\textit{early}) or spiral (\textit{late}), it is expected that the presence of strong arms in the latter may lead to an increase in any nonlensing flexion-like signal, or \textit{intrinsic flexion}, for a sample of galaxies.

Here and throughout, we describe the characteristic size in terms of the quadrupole image moments:

\begin{equation}\label{eq:a}
a = \sqrt{|Q_{11} + Q_{22}|}.
\end{equation}

\noindent The combination of a galaxy's  size, $a$, and measured flexion produces a scale-invariant \textit{dimensionless flexion} of a lensed object, $|a\vec{\f}|$. The same apparent galaxy image produced at different distances will have the same combination of $|a\vec{\f}|$. This dimensionless flexion becomes an excellent measure of the intrinsic flexion in a distribution of galaxies, with the scatter in that distribution producing a measure of the ``noise'' in the measured flexion for an object of a given size \citep{Goldberg/Bacon,Goldberg/Leonard}. 

The vector components of  $|a\vec{\f}|$ are treated as individual flexion measures with a mean of zero and added in quadrature. If the dimensionless flexion in one dimension is truly normally distributed, it is expected that the standard deviation of a bivariate Gaussian follows a relation $\langle a \cdot\f \rangle = \sqrt{\frac{\pi}{2}} \sigma_{a\cdot\f}$. This relation can be helpful in comparing to other studies that may occasionally use a metric on the average flexion.

Understanding the effect of shape, color and distance on the measured dimensionless flexion will then provide insight into reducing the scatter in this value and increasing the useful flexion signals in future studies.
\section{Lenser}\label{sec:Lenser}

\subsection{\texttt{Lenser} Pipeline}\label{sec:Lenser_Pipeline}

One factor limiting research into gravitational lensing flexion signals has been the lack of a robust analysis tool.  As such, we have developed \texttt{Lenser}\footnote{https://github.com/DrexelLenser/Lenser} -- a fast, open-source, minimal-dependency Python tool for estimating lensing signals from real survey data or realistically simulated images.  The module forward-models second-order lensing effects, performs a point spread function (PSF) convolution, and minimizes a parameter space.

Previous studies on flexion signals have made use of several techniques, including: moment analysis of light distribution \citep{Goldberg/Leonard,HOLICs}, decomposing images into ``shaplet'' basis sets \citep{Goldberg/Bacon,Goldberg/Leonard,MasseyFlexion}, and exploring the local potential field through a forward-modeling, parameterized ray tracing known as Analytic Image Modeling (AIM) \citep{Flexion/Cain}.  \texttt{Lenser} is intended as a hybrid approach, first using a moment analysis to localize a best-fit lensing model in parameter space and then performing a local minimization on the model parameters (seven lensing potential parameters, six shape parameters).

The unlensed intensity profile of a galaxy can be well described by a particular model with a corresponding set of model parameters \citep{SersicA,Graham/Driver}.  We employ a modified S\'{e}rsic-type intensity profile for the modeling galaxies:

\begin{equation}\label{eq:SerProf}
I(\theta)=I_0\exp\left[-\left(\frac{\theta'}{\theta_s}\right)^{1/n_s}\right],
\end{equation}

\noindent where $I_{0}$ is the central brightness, $\theta_{s}$ is the characteristic radius, $n_{s}$ is the S\'{e}rsic index (a measure of curve steepness), and the radial coordinate $\theta'$ is given by

\begin{equation}\label{eq:thetap}
\theta' = \sqrt{(x/q)^2+y^2},
\end{equation}

\noindent where $x$ and $y$ are the centroid-subtracted source-plane coordinates rotated appropriately by an orientation angle $\phi$, and $q$ is the semimajor-to-semiminor axis ratio of the galaxy. \footnote{$\theta'=\theta$ therefore corresponds to a circularly symmetric galaxy in the limit of no lensing.}

Equations \ref{eq:lensTaylor}, \ref{eq:SerProf}, as well as the centroid ($\theta_0^1,\theta_0^2$), $q$, and $\phi$, create a thirteen-parameter space to describe a galaxy.  Recognizing the existence of the shear/ellipticity degeneracy, we initially set shear to zero ($\psi,_{ij}=0$) and absorb the degenerate parameters into the intrinsic ellipticity described by $q$ and $\phi$.  In the context of smoothed mass mapping, the inferred shear can be used as a prior. This leaves us with a ten-parameter space given by

\[p=\left\{\theta_0^1,\theta_0^2,n_s,\theta_s,q,\phi,\psi,_{111},\psi,_{112},\psi,_{122},\psi,_{222}\right\}.\]

The first stage of \texttt{Lenser} uses an input galaxy image to estimate and subtract a background and estimate the sky and Poisson noise to use as a noise map weighting. If relevant noise maps are available, there is also an option to utilize those directly. A mask is then added so as to include only relevant pixels in the input image, reducing error from spurious light sources.  This elliptical mask is taken to be the region of the postage stamp where ${\rm S/N}>2.5$.  The second stage of \texttt{Lenser} estimates brightness moments from an unweighted quadrupole and hexadecapole calculation to be used as an initial guess for the galaxy model.  Shape and flexion parameters are estimated from image moments as in \citet{Bartelmann/Schneider} and \citet{Goldberg/Leonard}, respectively.   At this stage, initial guesses are provided for the entire parameter space except for $\{n_s, \theta_s\}$.

With initialized parameter estimates provided by the measured image moments, the final stage of the \texttt{Lenser} pipeline employs a two-step $\chi^2$ minimization:
\begin{enumerate}%[label=(\roman*)]
	\item first minimizing over the initially coupled subspace $\{n_s, \theta_s\}$
	\item a final full ten-parameter space local minimization.
\end{enumerate}

The AIM portion of the hybrid \texttt{Lenser} method (AIM-L) builds on the original AIM method of \citet{Flexion/Cain} (AIM-C) in two major ways.  First, while AIM-C uses an elliptical Gaussian intensity profile for the modeling galaxies (i.e. $n_s$ is fixed to $0.5$), AIM-L uses the modified S\'{e}rsic-type profile of Eq. \ref{eq:SerProf}.  \citet{Flexion/Cain} find that when a S\'{e}rsic profile is used, the modeling is not robust due to degeneracies between the image brightness normalization, image size, and $n_s$, as well as the parameter space simply being too large.  The authors decide to only model the lensing distortions of the galaxy isophotes and accurately fit the flexion at the cost of poorly fitting the image normalization and image size, and not fitting $n_s$ at all.  In AIM-L, we fit $n_s$, but do not fit the image normalization.  AIM-L therefore maintains the same size parameter space as AIM-C.  Second, AIM-L uses a two-step minimization approach, as described above, whereas AIM-C only performs a single local minimization.  We find that we are able to decouple $n_s$ and $\theta_s$ from each other by using the following subspace minimization method: we take the estimate of $q$ from the image moments portion of the hybrid \texttt{Lenser} method, iterate over the range of reasonable $n_s$ values for galaxies, and make use of the relation

\begin{equation}
\theta_s = a\sqrt{\frac{2}{1+q^2}\frac{\Gamma(2 n_s)}{\Gamma(4 n_s)}},
\end{equation}

\noindent where $a$ is given by Eq.\ref{eq:a} and $\Gamma(x)$ is the Gamma function, in order to calculate $\theta_s$ at each iteration.  This procedure provides estimates for $\{n_s, \theta_s\}$ before proceeding to the full local minimization.

With AIM-L, we find that we are able to robustly fit our entire parameter space.  This allows us to use fit $n_s$ values as a simple way to classify galaxies by type, which is useful for flexion-based studies (see Sec. \ref{sec:Results}).

\subsection{Covariance Testing}\label{sec:Covariance_Testing}

Since \texttt{Lenser} is a forward-modeling code, the user can specify a set of input parameters and create an image of a lensed galaxy.  It is, therefore,  possible to use \texttt{Lenser} in order to compute a covariance matrix for our parameter space by simulating an ensemble of postage stamp images (a "stamp collection") with known input parameters, $\hat{p}_i$, and noise, and then running each of the postage stamps through \texttt{Lenser} for fitting. To test the response of \texttt{Lenser} to noise, each postage stamp has identical input parameters and noise maps, but additional, unique Gaussian noise injected into each. The covariance matrix is given by $\sigma_{ij}^2=\langle (p_i-\langle p_i \rangle)(p_j-\langle p_j \rangle)\rangle$, where $p_i$ are the reconstructed parameters.  Once the covariance matrix is calculated, we are able to compute the marginalized $1\sigma$ uncertainty on each parameter simply by taking the square root of the diagonal: $\sigma_i = \sqrt{\sigma_{ii}^2}$.

\begin{figure}
	\includegraphics[width=\columnwidth]{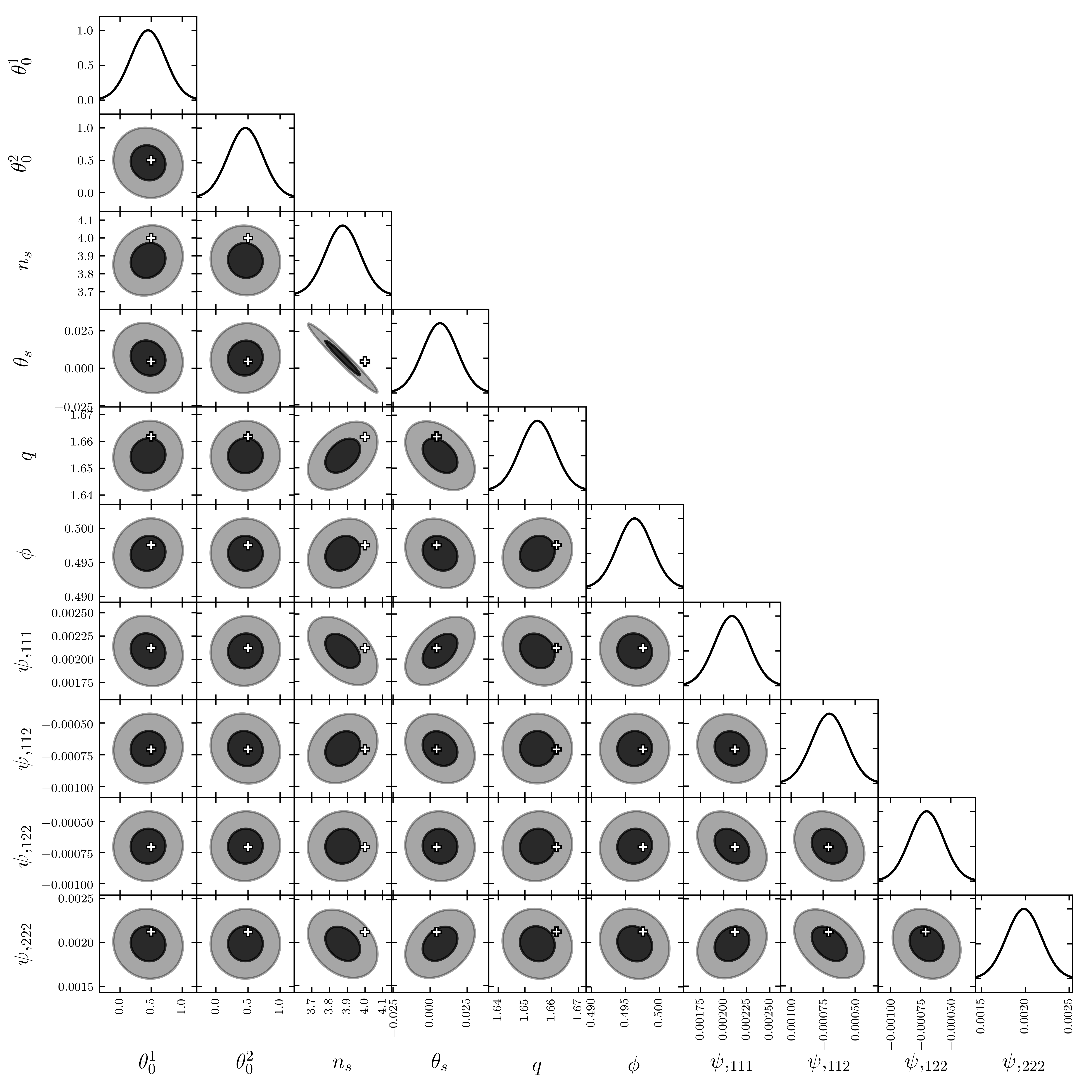}
	\caption{Covariance matrix for a stamp collection of 1000 images, with a noise map that matches that of the low-redshift catalog of Sec. \ref{sec:Catalog1}.  Here, we show $1\sigma$ and $2\sigma$ confidence ellipses below the diagonal (darker and lighter shades, respectively), $1\sigma$ Gaussians along the diagonal, and the locations of $(\hat{p}_i,\hat{p}_j)$ denoted by white plus signs. The galaxy in each postage stamp is chosen to be canonically elliptical ($n_s=4$). The galaxy also has realistic, nonzero ellipticity, shear, and flexion values. The centroid is dithered within a single pixel and hence is randomized for each image.  We note that, due to the shear/ellipticity degeneracy in the presence of nonzero input shear as we have here, $\hat{q}$ and $\hat{\phi}$ will not be reconstructed by \texttt{Lenser}.  Hence, the location of the white plus sign for these values is estimated analytically by adding the input intrinsic ellipticity and shear components together. The centroid and $\theta_s$ are in units of pixels, $\phi$ is in units of radians, $n_s$ and $q$ are dimensionless, and the $\psi,_{ijk}$ flexion terms have units of pixels$^{-1}$.}
	\label{fig:lensercov}
\end{figure}

Fig. \ref{fig:lensercov} shows the $1\sigma$ and $2\sigma$ confidence ellipses (and $1\sigma$ Gaussians along the diagonal) for the covariance matrix of a particular stamp collection. Additionally, since the input parameters are known, we can display them on top of the error ellipses to explicitly compare each $\hat{p}_i$ to $\langle p_i\rangle$. The white plus sign in each matrix element $(i,j)$ indicates the location of ($\hat{p}_i$,$\hat{p}_j)$. Successful fits will have white plus signs that fall within the error ellipses.  We clearly see from Fig. \ref{fig:lensercov} that \texttt{Lenser} is able to appropriately reconstruct the input parameter space. \footnote{For the set of parameters used in the covariance analysis of Fig. \ref{fig:lensercov}, we note that $(\hat{n_s},\hat{\theta_s})$ falls outside the error ellipse.  When using \texttt{Lenser} to create the images for the stamp collection, the user needs to specify additional input values outside of the parameter space, such as $I_0$ and $a$, where the latter is used to derive $\theta_s$ from $n_s$.  We attribute the discrepancy in $\hat{n_s}$ versus $\langle n_s \rangle$ to the degeneracies that exist between image brightness normalization, $n_s$, and image size that occur for large $n_s$/ $\theta_s \ll a$ (see Sec. \ref{sec:Lenser_Pipeline}). Despite this, since $I_0$ and $a$ are not in the parameter space, a robust fit is still achieved.} It is also evident that reasonable correlations exist in the parameter space.  For example, we see that $n_s$ and $\theta_s$ are anticorrelated, as expected.

\section{DATA CATALOGS}\label{sec:Data_Catalogs}

\subsection{Catalog 1: Low-Redshift Objects}\label{sec:Catalog1}

As the main drive for this study is to understand the effect of morphological shape on the measurement of flexion signal in source galaxies, a large catalog of high-quality galaxies was needed. A low-redshift catalog reduces the likelihood of any shape distortions from intervening fields, allowing for an improved estimation of the intrinsic flexion for statistical analysis of weakly lensed source objects.

The EFIGI (Extraction de Formes Idealisees de Galaxies en Imagerie) project was developed to robustly measure galaxy morphologies \citep{EFIGI}. Thus, the publicly available catalog of images and morphological classifications is well suited for our study on intrinsic flexion. The catalog includes imaging data for a total 4458 galaxies from the RC3 Catalogue, ranging from z = 0.001 to z = 0.09. The full catalog merges data from several standard surveys, pulling from Principal Galaxy Catalogue, Sloan Digital Sky Survey, Value-Added Galaxy Catalogue, HyperLeda, and the NASA Extragalactic Database. Imaging data used in the EFIGI study was obtained from the SDSS DR4 in the $u$ and $r$ bands, with accompanying PSF images. The camera system used a pixel scale of 0.396 arcsec, though the publicly available images were rescaled for a uniform 255 $\times$ 255 pixels$^2$ stamp for purposes of the original study. The varied sources for this large sample should reduce any strong effects from real lensing distortions on the catalog as a whole.

Galaxies were reduced from the more complex morphological classifications in EFIGI to a simplified \textit{early}/\textit{late}/\textit{irregular} scheme, with \textit{irregular} objects excluded completely for subsequent analysis. Objects were initially visually classified with a classifier value corresponding to a characteristic shape in the Hubble sequence (EHS), with a corresponding uncertainty. For increased confidence in the simplified scheme, the intermediate range objects were excised. The final catalog of objects to be used in the study was reduced to a total of 1551 high-quality, low-redshift objects (597 \textit{early}/954 \textit{late}).

\begin{figure}
	\includegraphics[width=\columnwidth]{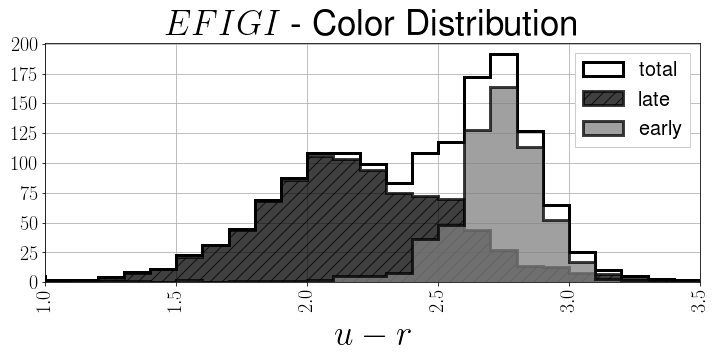}
	\caption{Color distribution of the 1551 analyzed EFIGI objects}
	\label{fig:EFIGI_ur}
\end{figure}

Galaxy color measurements in $u$ and $r$ bands from SDSS DR7 were used for optimal morphology separation \citep{Strateva_2001}. The interplay between color, shape and distance (source object redshift) should provide effective criteria for selecting source
galaxies. The final distribution of galaxy color measurements for Catalog 1 is shown in Fig. \ref{fig:EFIGI_ur}.

\subsection{Catalog 2: High-Redshift Objects}\label{sec:Catalog2}

An analysis of high-redshift objects is also useful for investigating how the morphology of real source galaxies affects the measured flexion and what selections should be applied to increase the signal-to-noise in future flexion-based studies.

The CANDELS \citep{CANDELS_2011} Program contains a large catalog of high-redshift ($z$ = $0.15$ to 8), deep-imaging galaxies using the HST WFC3/IR and ACS camera systems, with an emphasis on deep-probing faint, distant galaxies, with the added
effect of including many bright background objects that may be
analyzed for weak lensing effects. The HST camera systems operate at a pixel scale of  0.03 arcsec/pixel. Previous work has produced morphological classifications of
the observed galaxies in the COSMOS field across a \textit{early/late} scheme \citep{Tasca}.

\begin{figure}
	\includegraphics[width=\columnwidth]{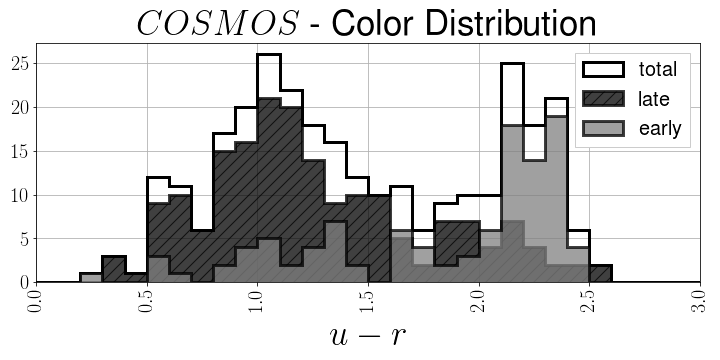}
	\caption{Color distribution of the 293 analyzed CANDELS-COSMOS objects}
	\label{fig:COSMOS_ur}
\end{figure}

The creation of image stamps for use in the \texttt{Lenser} pipeline entailed locating objects with known position, redshift, and morphology across several separate catalogs. Using a kd-tree method object matching technique \citep{KDtree} with previous redshift catalogs in the COSMOS field, 1492 initial galaxies of known morphological type, color and redshift were identified and analyzed. Several metrics were used to ensure that identified objects were large enough for analysis. 

In order to select only highest resolution objects, cuts were made in object size and signal strength to ensure the best-quality fits in high-$z$ objects. A hard cut of $|a\cdot\mathcal{F}| \leq$ 1 and $\chi^2 \leq$ 1.5 was applied in keeping with previous flexion studies\citep{Flexion/Cain}. This was undertaken to eliminate any concerns of background field biasing the analysis. A subset of 293 objects was achieved, with the same distribution of morphological type as the larger Catalog 1, to ensure a fair comparison in analyzed behavior (107 \textit{early}/186 \textit{late}). The final distribution of galaxy color measurements for Catalog 2 is shown in Fig. \ref{fig:COSMOS_ur}.

\section{Results}\label{sec:Results}
\subsection{Low-Redshift Objects}\label{sec:Results_Low-z}

The low-$z$ galaxy and lensing parameters, with uncertainties, were estimated using \texttt{Lenser} in the ``zero shear'' limit. In addition to the flexion estimate, the pipeline yields a characteristic size $a$ (Eq. \ref{eq:a}) as well as a S\'{e}rsic index, which allows an independent estimate of morphology. Canonically, $n_s$ = 1 for spiral galaxy profiles and $n_s$ = 4 for elliptical galaxy profiles, but there exists a wide variability.

\begin{figure}
	\includegraphics[width=3in]{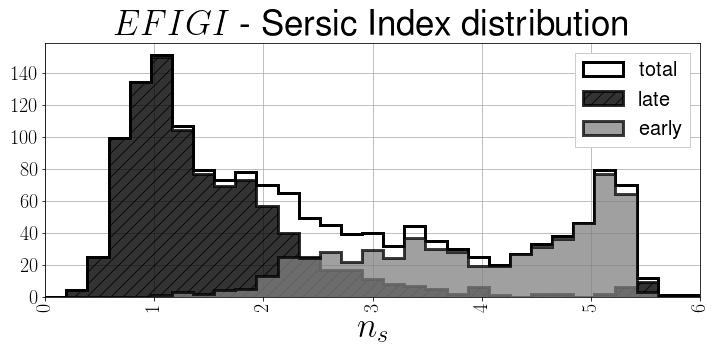}
	\caption{S\'{e}rsic index distribution of the analyzed low-$z$ objects.}
	\label{fig:EFIGI_ns}
\end{figure}

\begin{figure}
	\includegraphics[width=3in]{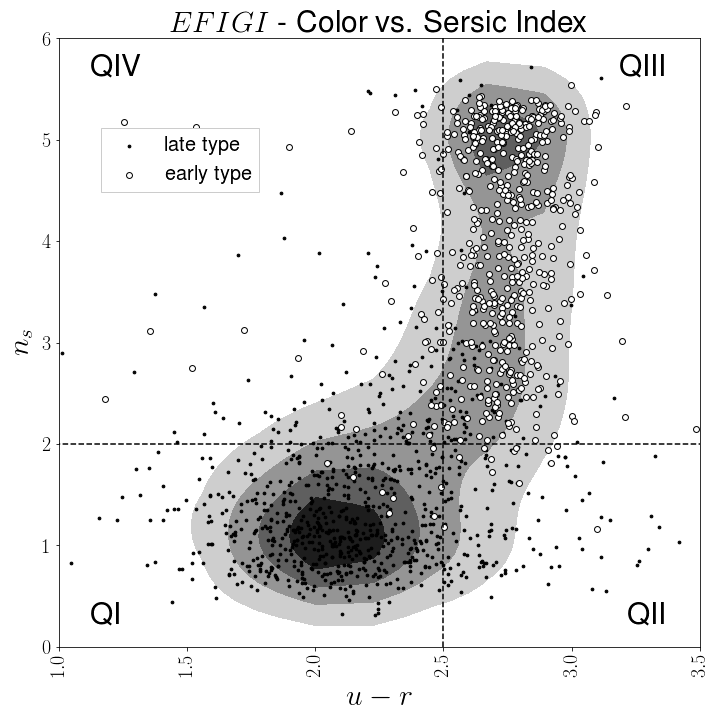}
	\caption{Color vs S\'{e}rsic index plot, split into major areas of interest}
	\label{fig:quad_plot1}
\end{figure}

The distribution of measured S\'{e}rsic
index and a comparison with color measures are shown in Fig. \ref{fig:EFIGI_ns} and Fig. \ref{fig:quad_plot1} respectively. The resulting scatter plot of Fig.  \ref{fig:quad_plot1} is overlayed on a two-dimensional kernel density estimate (KDE) for visualization purposes, and morphological type classifications are plotted separately.

%discussing clustering
Looking at Fig. \ref{fig:quad_plot1}, two distinct clusters in the data are readily apparent. There is a clearly separated grouping of objects in the upper-right locus, with a denser but wider grouping concentrated in the lower-left locus. This is consistent with the expectation that \textit{early}-type objects are redder, with a higher $n_s$ than \textit{late}-type objects. A morphological separation in the data mostly follows this trend, with \textit{early}-type objects almost exclusively clustered to the right (clearly corresponding to a ``redder'' measure of galaxy color), while the \textit{late}-type objects are more broadly spread but  largely concentrated to the lower left. 

\begin{figure}
	\includegraphics[width=3in]{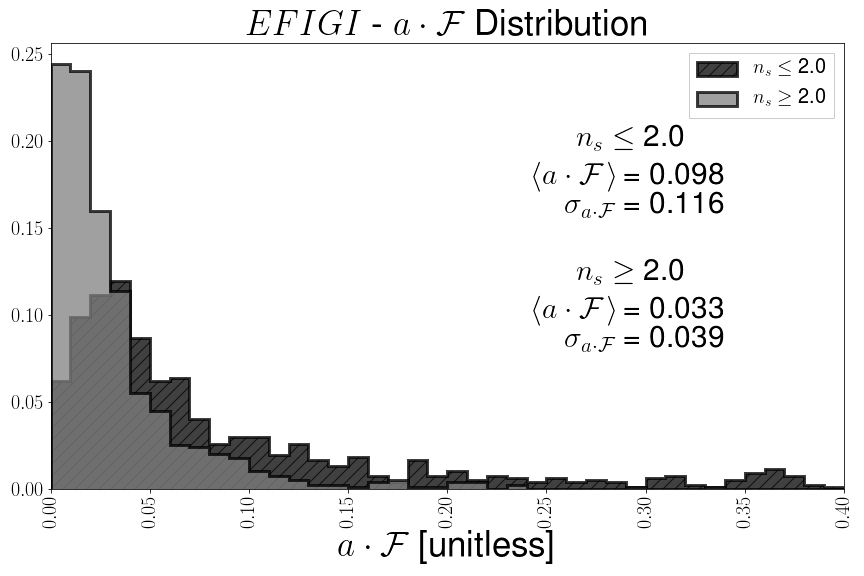}
	\caption{Distributions of $|a\vec{\f}|$ for low redshift galaxies, split by S\'{e}rsic index (782 above/769 below).}
	\label{fig:afsersic1}
\end{figure}

\begin{figure}
	\includegraphics[width=3in]{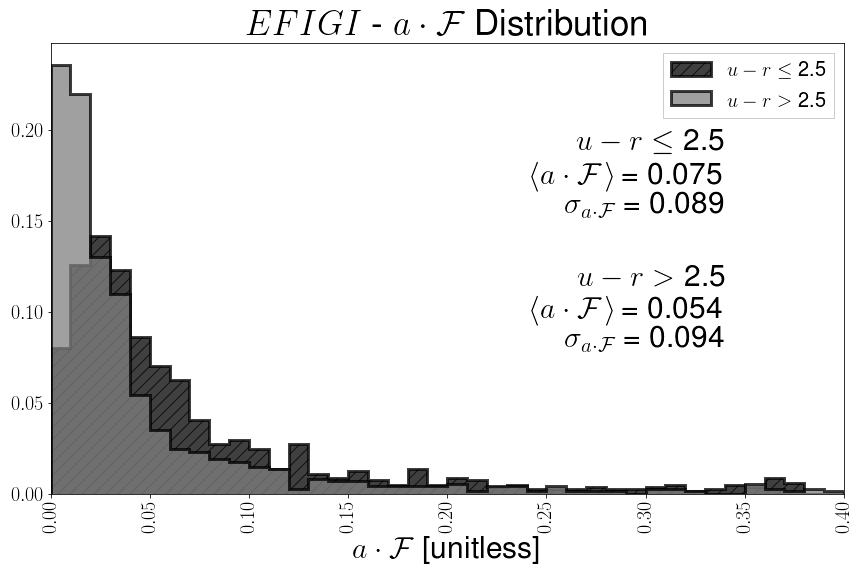}
	\caption{Distributions of $|a\vec{\f}|$ for low redshift galaxies, split by median $u-r$ (738 above/813 below).}
	\label{fig:afcolor1}
\end{figure}

We can view the behavior of dimensionless flexion along different separations in the data. Initial estimates suggest that the intrinsic flexion scatter is low, surprisingly dependent on Sérsic index (Fig. \ref{fig:afsersic1}) but independent of color (Fig. \ref{fig:afcolor1}). Here the choice of $n_s$ = 2 as a splitting value represents the approximate shifting point in the distribution of measured $n_s$ (Fig. \ref{fig:EFIGI_ns}). A choice of $n_s$ = 4 also marks an upper cut, which corresponds to a  de Vaucouleurs-esque profile for Eq. \ref{eq:SerProf}. The lower value was chosen to favor a lower intrinsic scatter, while not excluding useful measures.

Similarly, a natural cut in the color distribution can be seen around $u-r$ = 2.5 (Fig. \ref{fig:EFIGI_ur}). As expected, we anticipate \textit{early}-type galaxies should be more well behaved in a measure of intrinsic flexion, as the galaxy profiles are less likely to contain artifacts to distort a measurement. This behavior matches the split in measured S\'{e}rsic index. 

\begin{figure}
	\includegraphics[width=3in]{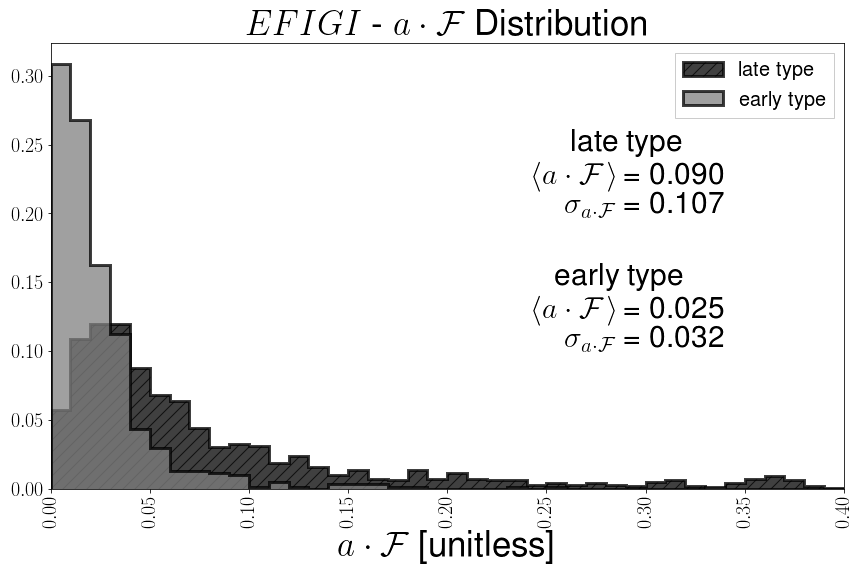}
	\caption{Distributions of $|a\vec{\f}|$ for low redshift galaxies, split by morphological type (597 \textit{early}/954 \textit{late}).}
	\label{fig:aftype1}
\end{figure}

More broadly, we can see the topography of the S\'{e}rsic index/color
diagram can be divided into
``quadrants'' (Fig. \ref{fig:quad_plot1}), based on the bimodal natures of both
color and measured $n_s$ values. Thus, we
expect large clustering in the upper-right and lower-left quadrants of
the diagram, which is indeed evident. As such, the majority of objects in QI are expected to be \textit{late}-type objects, and conversely, QIII are \textit{early}-type objects. There exists a more mixed clustering in objects in QII, indicating that this range of profile and color measures is where the transition between a classically elliptical galaxy and those with more spiral-like features, containing the young massive stars that dominate color measurement. The fourth quadrant is almost completely void of measurements, indicating little overlap between a lower color and broader profile.

Using only these hard cuts on the $n_{s}$ and $u-r$ values, we can see (Table \ref{table:cat1}) how the intrinsic scatter in dimensionless flexion can be greatly reduced:

%----- TABLE CATALOG 1 --------
\begin{table}
	\label{table:cat1}
	\centering
	\begin{tabular}{||l||c||r||} 
		\hline
		\textbf{Quadrant} & \textbf{$\sigma_{|a\cdot\f|}$} & \textbf{Sample Size}\\ 
		\hline
		QI                & 0.0958
		& 631                                                   \\
		QII               & 0.1664 & 138                                                          \\
		QIII              & 0.0336 & 600                                                          \\
		QIV               & 0.0505 & 182                                                          \\
		\hline
		Full Set		  &
		0.0920 & 1551															\\
		\hline
	\end{tabular}
	\caption{Measured scatter in $|a\vec{\f}|$ for Catalog 1}
\end{table}

\subsection{High-Redshift Objects}\label{sec:Results_High-z}

Following the same procedure as Catalog 1, the objects in Catalog 2 were analyzed using the \texttt{Lenser} pipeline. As in the more robust low-redshift study, a comparison of color measurements and the measured $n_s$ values shows distinctive split in the data. Object measurements are clustered in either lower $n_s$/less red or higher $n_s$/more red subgroups. 

\begin{figure}
	\includegraphics[width=3in]{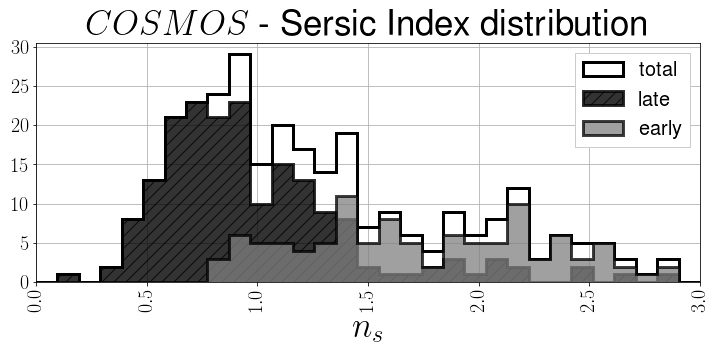}
	\caption{S\'{e}rsic index distribution of the analyzed high-$z$ objects}
	\label{fig:COSMOS_ns}
\end{figure}

\begin{figure}
	\includegraphics[width=3in]{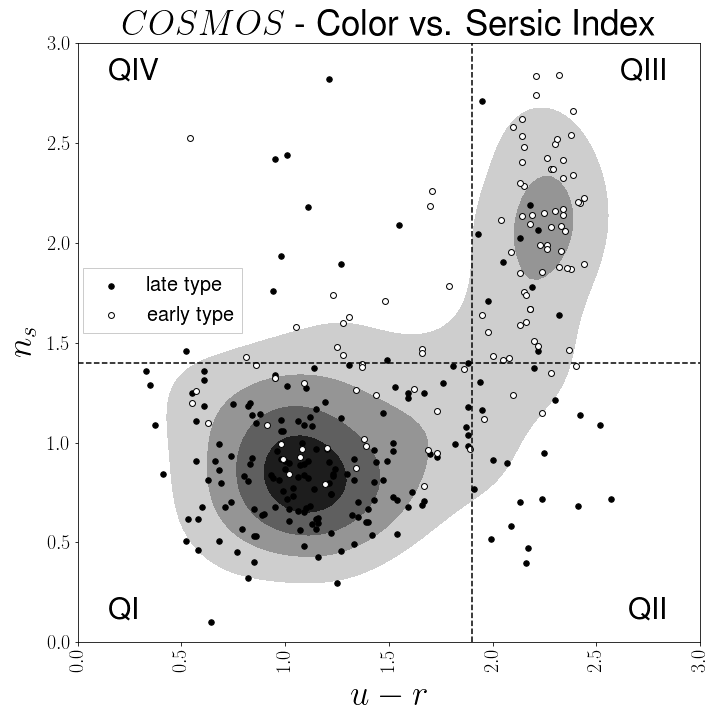}
	\caption{Color vs S\'{e}rsic index plot, high-$z$ catalog}
	\label{fig:tri_plot2}
\end{figure}

\begin{figure}
	\includegraphics[width=3in]{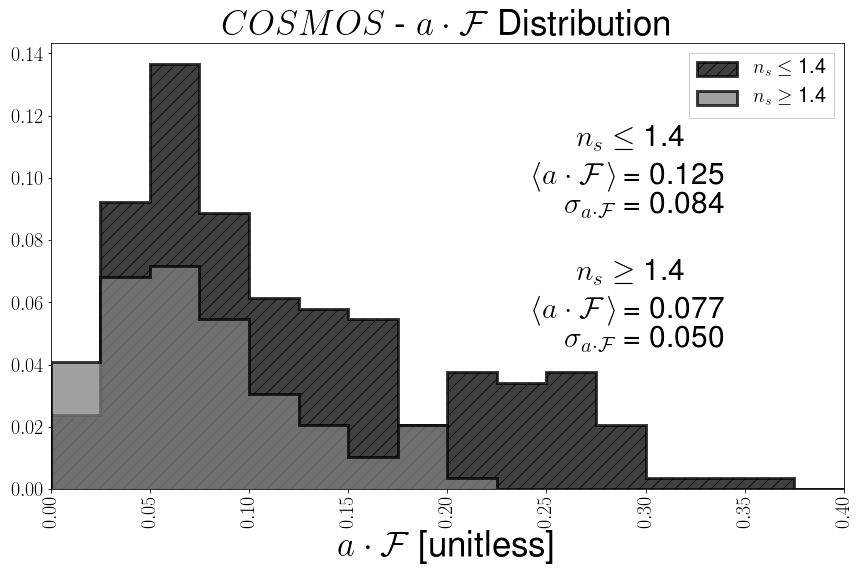}
	\caption{Distributions of $|a\vec{\f}|$ for high-$z$ objects, split by S\'{e}rsic index (94 above/199 below).}
	\label{fig:afsersic2}
\end{figure}

\begin{figure}
	\includegraphics[width=3in]{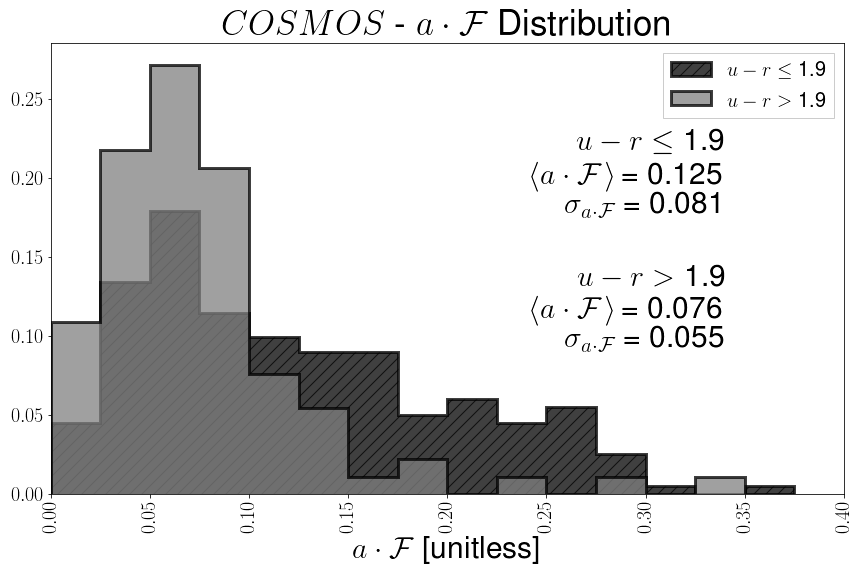}
	\caption{Distributions of $|a\vec{\f}|$ for high-$z$ objects, split by  $u-r$ (92 above/201 below).}
	\label{fig:afcolor2}
\end{figure}

\begin{figure}
	\includegraphics[width=3in]{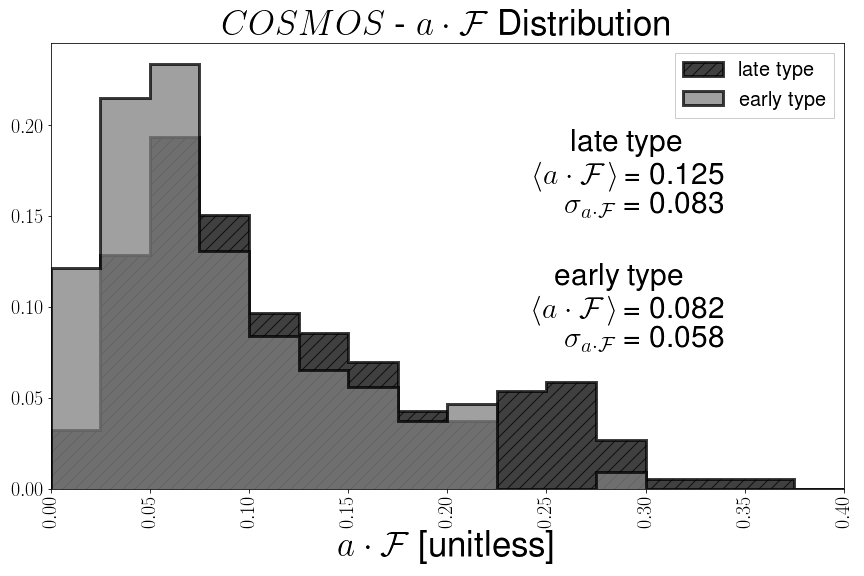}
	\caption{Distributions of $|a\vec{\f}|$ for high-$z$ objects, split by morphological type (107 \textit{early}/186 \textit{late}).}
	\label{fig:aftype2}
\end{figure}

Here, the split in the S\'{e}rsic index distribution is taken to be $n_s$ = 1.4, as there is a noticeable drop in the distribution shape (Fig. \ref{fig:COSMOS_ns}). A similar clustering by morphological type is seen in the color-Sérsic index comparison (Fig. \ref{fig:tri_plot2}). Measures of the dimensionless flexion distributions for splits in $n_s$, color and independently identified morphology are shown (Fig. \ref{fig:afsersic2}, \ref{fig:afcolor2}, \ref{fig:aftype2}). Note that for this sample of high-$z$ objects, all major splits in measured galaxy qualities produce a lower intrinsic scatter.

Again, we see that by using broad cuts for strategic points in the full data set, the expected noise in $|a\vec{\f}|$ can be consistently and systematically reduced. Though the separation in clusters is less, owing to a smaller range in measured S\'{e}rsic index values, a desirable reduction in the expected noise is achievable, just as in the low-redshift objects.

%%----- TABLE CATALOG 2 (PSF)--------
\begin{table}
	\centering
	\begin{tabular}{||l||c||r||} 
		\hline
		\textbf{Quadrant} & \textbf{$\sigma_{|a\cdot\f|}$} & \textbf{Sample Size} \\ 
		\hline
		QI                & 0.0832 & 176                                                          \\
		QII               & 0.0856 & 23                                                          \\
		QIII              & 0.0370 & 69                                                          \\
		QIV               & 0.0665 & 25                                                          \\
		\hline
		Full Set		  &
		0.0777 & 293															\\
		\hline
	\end{tabular}
	\caption{Measured scatter in $|a\vec{\f}|$ for Catalog 2}
\end{table}

\section{Summary}\label{sec:Summary}

For both the high- and low-redshift catalogs, it can be seen that the observed dimensionless flexion distribution favors a smaller scatter in objects that are consistent with an \textit{early}-type shape (QIII in our shorthand). This behavior agrees with the expectation of galaxy shape providing a significant influence on measurable flexion signal. For high-redshift objects, which are cosmologically younger, we can still achieve a significant reduction in the intrinsic scatter for a sample of galaxies.

The major takeaway of this study is to show how detected flexion signal can be boosted through a selection discrimination on chosen source galaxies. By choosing identified source objects with favorable characteristics toward more \textit{early}-type galaxies, the anticipated noise in future lensing studies attributed to purely unlensed shape information can be reduced by a factor of more than two.  Selected source objects are expected to display a flexion noise drawn from the high color/high $n_s$ subset distribution, and scaled by the source galaxies measured size. 

As an example of this effect, a flexion-based detection for an elliptical source object of size $a \approx$ 0.5'' should be detectable with a S/N of 1 at a separation of 6'' from a $\sigma_v$ = 300 km/s lens galaxy. A similarly size late-type galaxy would need to be found at a separation of at most 3'', a less likely scenario. By selecting for more elliptical or \textit{early}-type objects future flexion based studies can boost signal analysis by properly weighting measures via where sources fall in the color and $n_s$ distributions, leading to improved constraints on the underlying local density. Furthermore, the extended-AIM approach used in \texttt{Lenser} is faster and more robust under pixel noise than previous approaches and will hopefully allow for increased use of flexion analysis in the lensing community. 

\section*{Acknowledgements}

%EFIGI
This work makes use of the \textit{EFIGI} catalog, which in turn made use of the Sloan Digital Sky Survey. Funding for the SDSS and SDSS-II has been provided by the Alfred P. Sloan Foundation, the Participating Institutions, the National Science Foundation, the U.S. Department of  Energy,  the  National  Aeronautics  and  Space  Administration,  the  Japanese Monbukagakusho, the Max Planck Society, and the Higher Education Funding Council for England. The SDSS Web Site is http://www.sdss.org/. 

%CANDELS COSMOS
This work is also based on observations taken by the CANDELS Multi-Cycle Treasury Program with the NASA/ESA HST, which is operated by the Association of Universities for Research in Astronomy, Inc., under NASA contract NAS5-26555. 

\section*{Data Availability}

The data underlying this article are available at https://github.com/DrexelLenser/Lenser. The datasets were derived from publicly available sources: (EFIGI) https://www.astromatic.net/projects/efigi; (COSMOS) https://irsa.ipac.caltech.edu/data/COSMOS/images/candels/ 

%%%%%%%%%%%%%%%%%%%%%%%%%%%%%%%%%%%%%%%%%%%%%%%%%%



\bibliographystyle{mnras}
\bibliography{lensing_MNRAS} % if your bibtex file is called example.bib

%\begin{thebibliography}{99}
%\end{thebibliography}

%%%%%%%%%%%%%%%%%%%%%%%%%%%%%%%%%%%%%%%%%%%%%%%%%%

%%%%%%%%%%%%%%%%% APPENDICES %%%%%%%%%%%%%%%%%%%%%

%%%%%%%%%%%%%%%%%%%%%%%%%%%%%%%%%%%%%%%%%%%%%%%%%%

% Don't change these lines
\bsp	% typesetting comment
\label{lastpage}
\end{document}